\documentclass[3p,times,twocolumn]{elsarticle}

\usepackage{ecrc}

\volume{00}

\firstpage{1}

\journalname{Nuclear Physics B Proceedings Supplement}

\runauth{}

\jid{nuphbp}

\jnltitlelogo{Nuclear Physics B Proceedings Supplement}

\usepackage{amssymb}

\usepackage[figuresright]{rotating}

\begin{document}

\begin{frontmatter}



\dochead{}


\title{Halo-Independent analysis of direct dark matter detection data for any WIMP interaction\tnoteref{label1}}
\tnotetext[label1]{Talk given at the 37th. International Conference on High Energy Physics (ICHEP 2014), July 2 to 9, 2014,  Valencia, Spain.}


 \author{Graciela  B Gelmini}

\address{Physics and Astronomy Department, UCLA}
\ead{gelmini@physics.ucla.edu}

\begin{abstract}
The halo independent comparison of direct dark matter detection data eliminates the need to make any assumption on the uncertain local dark matter distribution and is complementary to the usual data comparison which required assuming a dark halo model for our galaxy. The method, initially proposed for WIMPs with  spin-independent contact interactions, has  been generalized to any other interaction and applied to recent data on ``Light WIMPs".
\end{abstract}

%
%

\end{frontmatter}



Determining what the dark matter (DM), the most abundant form of matter in the Universe, consists of is one of the most fundamental open questions in physics and cosmology. Weakly interacting massive particles (WIMPs) are among the most experimentally sought after candidates.  Direct detection experiments attempt to observed the energy deposited within a detector by DM particles in the dark halo of our galaxy passing through it and colliding with nuclei in the detector. Three direct detection experiments, DAMA/LIBRA~\cite{Bernabei:2010mq}, 
CoGeNT~\cite{Aalseth:2010vx} and CDMS-II-Si~\cite{Agnese:2013rvf} 
have at present claims of  having observed potential signals of WIMP DM. The CRESST-II collaboration, with an upgraded detector does not find  any longer an excess in their rate attributable to a DM signal~\cite{Angloher:2014myn}, as they had found in their previous  2010 results~\cite{Angloher:2011uu}. 
All other direct detection searches, LUX, XENON100, XENON10, CDMS-II-Ge, CDMSlite, SuperCDMS, SIMPLE etc have produced only upper bounds on the interaction rate and annual modulation amplitude of a potential  WIMP signal (see an updated list of references in  Ref.~\cite{DelNobile:2014sja}). It is thus essential to compare these data to decide if the potential signals are compatible with each other and with the upper bounds of direct detection searches with negative results for any particular DM candidate.

The rate observed in a particular detector due to DM particles in the dark halo of our galaxy depends on three main elements: 1) the detector response to potential WIMP collisions within it, 2) the WIMP-nucleus cross section and WIMP mass and 3) the local density $\rho$ and velocity distribution $f(\vec{\rm v}, t)$ of WIMPs passing through the detector. The last element depends on the halo model adopted, which has considerable uncertainty. The usual {\bf{Halo-Dependent data comparison method}} fixes the three mentioned  elements of the rate,  usually assuming the Standard Halo Model (SHM) for the galactic halo, except for the WIMP mass $m$ and a reference  cross section parameter $\sigma_{\rm ref}$ extracted from the cross section, and data are plotted in the $(m, \sigma_{\rm ref})$ parameter space. For the usual spin-independent  (SI) interactions the reference cross section parameter is chosen to be the WIMP-proton cross section $\sigma_p$.
\begin{equation}
\label{SI}
\hspace{-20pt} \frac{d \sigma_T}{d E_R} = \sigma_p \frac{\mu_T^2}{\mu_p^2} \left| Z_T + (A_T - Z_T) \frac{f_n}{f_p} \right|^2 F_{{\rm SI}, T}^2(E_R)  \frac{m_T}{2 \mu_T^2 {\rm v}^2} .
\end{equation}
Here $E_R$ is the nuclear recoil energy, $Z_T$, $A_T$ and $m_T$ are respectively the atomic number, mass number and mass of the target nuclide $T$, $F_{{\rm SI}, T}(E_R)$ is the nuclear spin-independent form factor, $f_n$ and $f_p$ are the effective DM couplings to neutrons and protons, respectively,  and 
$\mu_T$ and $\mu_p$ are the WIMP-nucleus and the WIMP-proton reduced masses.

In the {\bf{Halo-Independent data comparison method}} one fixes the elements 1) and 2) of the rate, again except for a reference cross section parameter $\sigma_{\rm ref}$ extracted from the cross section,
but does not make any assumption about the element 3), circumventing in this  manner the uncertainties in our knowledge of the local characteristics of the dark halo of our galaxy~\cite{Fox:2010bz,
 Frandsen:2011gi, Gondolo:2012rs, Frandsen:2013cna, DelNobile:2013cta, HerreroGarcia:2011aa, HerreroGarcia:2012fu, Bozorgnia:2013hsa, DelNobile:2013cva,  DelNobile:2013gba, DelNobile:2014eta, Fox:2014kua, Gelmini:2014psa, Scopel:2014kba, DelNobile:2014sja, Feldstein:2014ufa, Bozorgnia:2014gsa}.
The main idea of this method is that the interaction rate at one particular recoil energy $E_R$ depends for any experiment on one and the same  function $\rho\eta( {\rm v}_{min}, t)/m$  (incorporated into the definition of 
$\tilde{\eta}({\rm v}_{min}, t)$ in Eq.~(3)) of the minimum speed ${\rm v}_{min}$ required for the incoming DM particle to cause a nuclear recoil with energy $E_R$. The function $\eta({\rm v}_{min})$  depends only on the local characteristics of the dark  halo of our galaxy.  Thus, all rate measurements  and bounds can be translated into measurements of and bounds on the unique function $\tilde{\eta}({\rm v}_{min}, t)$. This method was initially developed for SI  WIMP-nucleus interaction and only in Ref.~\cite{DelNobile:2013cva} extended to any other type of WIMP-nucleus interactions

It is easy to see that when computing the recoil spectrum,
\begin{equation}
\label{dRdER}
\frac{d R_T}{d E_R} = \frac{\rho}{m} \frac{C_T}{m_T} \int_{{\rm v} \geqslant {\rm v}_{min}(E_R)} \hspace{-24pt} d^3 {\rm v} \, f(\vec{\rm v}, t) \, {\rm v} \, \frac{d \sigma_T}{d E_R}(E_R, \vec{\rm v}) ,
\end{equation}
with the SI cross section in Eq.~1 the whole dependence on the local WIMP velocity distribution is contained in the function $\tilde{\eta}({\rm v}_{min}, t)$ (recall $\sigma_{\rm ref}=\sigma_p$ for SI interactions) 
\begin{equation}
\label{eta0}
\hspace{-22pt} \tilde{\eta}({\rm v}_{min}, t) \equiv   \frac{\rho \sigma_{\rm ref}}{m} \eta({\rm v}_{min}, t) \equiv \frac{\rho \sigma_{\rm ref}}{m} \int_{{\rm v} \geqslant {\rm v}_{min}}  \hspace{-0.2truecm} d^3 {\rm v} \, \frac{f(\vec{\rm v}, t)}{\rm v} .
\end{equation}
Due to the revolution of the Earth around the Sun,  the velocity integral  $\tilde{\eta}({\rm v}_{min},t)$ has an annual modulation generally well approximated by the first terms of a harmonic series,
\begin{equation}
\label{etat}
\tilde{\eta}({\rm v}_{min}, t) \simeq \tilde{\eta}^0({\rm v}_{min}) + \tilde{\eta}^1({\rm v}_{min}) \cos\!\left[ \omega (t - t_0) \right] ,
\end{equation}
where $t_0$ is the time of the maximum of the signal and $\omega = 2 \pi/$yr. The time average unmodulated and the modulated components $\tilde{\eta}^0$ and $\tilde{\eta}^1$ enter respectively in the definition of  the unmodulated and modulated parts of the rate.

For a particular WIMP candidate $\tilde{\eta}({\rm v}_{min}, t)$ must be common to all experiments.  Measurements and upper bounds on the time averaged rate and  the annual modulation amplitude of the rate can be mapped onto the $({\rm v}_{min}, \tilde{\eta})$ plane. By $\tilde{\eta}$ we understand either $\tilde{\eta}^0$ or $\tilde{\eta}^1$. To be compatible all experiments must measure the same functions $\tilde{\eta}^0$ or $\tilde{\eta}^1$ of ${\rm v}_{min}$.

The difficulty we want to address is how to do the same, i.e. compare direct detection data in the $({\rm v}_{min}, \tilde{\eta})$ plane, when the
differential scattering cross section does not have a simple $1/{\rm v}^2$ dependence on the speed ${\rm v}$ of the DM particle, but a more general dependence, such as two terms with different dependence on the speed ${\rm v}$. Consider for example  a fermionic WIMP interacting with the nucleus via a magnetic dipole moment $\lambda_\chi$, the so-called ``magnetic-dipole dark matter" (MDM), ${L}_{\rm int} = ({\lambda_\chi}/{2}) \, \bar\chi \sigma_{\mu \nu} \chi F^{\mu\nu}$ which leads to the cross section~\cite{Sigurdson:2004zp},
 \begin{eqnarray}
\label{MDMsigma}
\hspace{-20pt} \frac{d \sigma_T}{d E_R} & = &
\alpha \lambda_\chi^2 \left\{ Z_T^2 \frac{m_T}{2 \mu_T^2} \left[\frac{1}{{\rm v}_{min}^2} - \frac{1}{{\rm v}^2} \left( 1 - \frac{\mu_T^2}{m^2} \right) \right]  \right.
 \nonumber\\
& \times & \left. F_{{\rm SI}, T}^2(E_R({\rm v}_{min})) \right. 
 \nonumber\\
 & + & \left. \frac{\hat\lambda_T^2}{{\rm v}^2} \frac{m_T}{m_p^2} \left( \frac{S_T+1}{3S_T} \right) F_{{\rm M}, T}^2(E_R({\rm v}_{min})) \right\} .
\end{eqnarray}
Here $\alpha = e^2 / 4 \pi$ is the electromagnetic fine structure constant, $m_p$ is the proton mass, $S_T$ is the spin of the target nucleus, and $\hat\lambda_T$ is the magnetic moment of the target nucleus in units of the nuclear magneton $ e / (2 m_p)= 0.16$ GeV$^{-1}$.  The definition of the reference cross section parameter  for this cross section is arbitrary; a possible choice is $\sigma_{\rm ref} \equiv \alpha \lambda_\chi^2$ (the plots for MDM that follow use this definition). The first term corresponds to the dipole-nuclear charge coupling, and the corresponding  charge form factor coincides  with  the usual spin-independent nuclear form factor $F_{{\rm SI}, T}(E_R)$.  This is usually taken to be the Helm form factor~\cite{Helm:1956zz} normalized to $F_{{\rm SI}, T}(0)=1$. The second  term, corresponds to the coupling of the DM magnetic dipole to the magnetic field of the nucleus, and the corresponding nuclear form factor is the nuclear magnetic form factor $F_{{\rm M}, T}(E_R) $. This magnetic form factor includes the contributions of  the magnetic currents due to the orbital motion of the nucleons and of  the intrinsic nucleon magnetic moments (proportional to the spins).

\begin{figure}[t]
\centering
\includegraphics[width=0.45\textwidth]{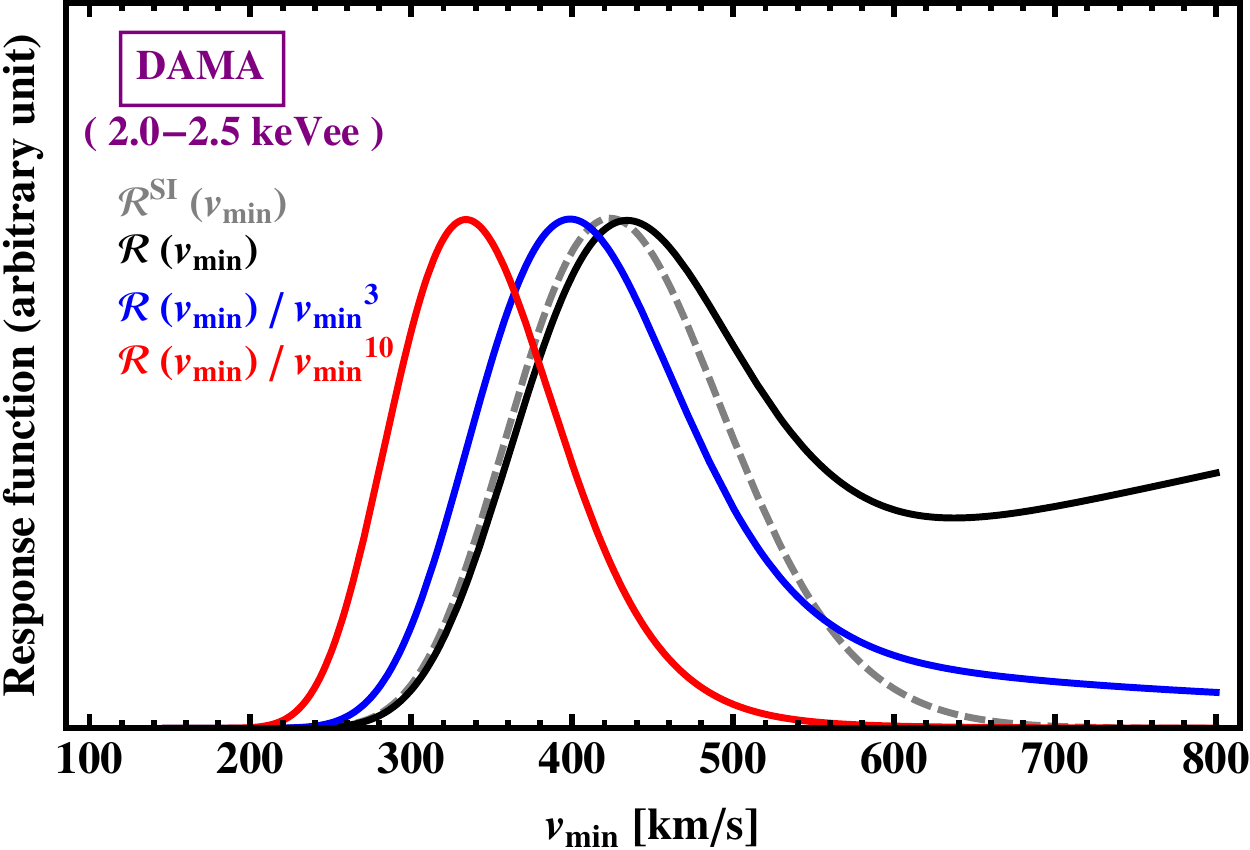}
\label{responsefunction}

\vspace{-0.3cm}

\caption{
Example of response functions  $\mathcal{R}_{[E'_1, E'_2]}({\rm v}_{min})$  for Si interactions  (gray-dashed line) and   for MDM   with arbitrary normalization, for the 2.0 to 2.5 keVee detected energy interval  of DAMA /LIBRA, assuming interaction with Na.  For MDM the function is regularized  by multiplying it by ${\rm v}_{min}^{-r}$, with $r$ a positive integer. The best choice is $r=10$ (red continuous line).  See 
Ref.~\cite{DelNobile:2013cva} for details.}
\end{figure}

Notice that the cross section in Eq.~\ref{MDMsigma} contains two  terms with different dependences on the DM particle speed ${\rm v}$. When these terms are integrated over the velocity distribution to find the interaction rate, instead of a unique function $\tilde{\eta}({\rm v}_{min})$, each term has its own function of ${\rm v}_{min}$ multiplied by its own detector dependent coefficient. It seems thus impossible  to translate a rate measurement or bound into only one of the two ${\rm v}_{min}$ functions contributing to the rate. In other cases, such as  that of ``Resonant DM"~\cite{Pospelov:2008qx},  the cross section has an energy dependence with a shape that depends on the target nucleus. Thus each target has its own function of ${\rm v}_{min}$, and again it seems impossible to find one and the same common function analogous to $\tilde{\eta}({\rm v}_{min})$ so that all rate measurements and bounds can be mapped onto it.  Following Ref.~\cite{DelNobile:2013cva}, we show here how this difficulty can be circumvented and encode for general interactions all the halo dependences of the observable rate again in the sole function $\tilde{\eta}({\rm v}_{min})$.

The differential recoil rate is not directly experimentally accessible because of energy dependent efficiencies and 
energy resolutions functions and because what is often measured is a part $E'$ of the recoil energy $E_R$. The observable differential rate is
\begin{equation}
\label{Obs-rate}
\hspace{-5pt} \frac{dR}{dE'} = \epsilon(E') \, \int_0^\infty dE_R \, \sum_{T} C_{T} \, G_{T}(E_R,E') \, \frac{dR_{T}}{dE_R} ,
\end{equation}
where $E'$ is the detected energy, often quoted in keVee (keV electron-equivalent) or in photoelectrons and
 $\epsilon(E')$ is  a counting efficiency or cut acceptance. $G_T(E_R, E')$  in a (target nuclide and detector dependent) effective energy resolution function that gives the probability that a recoil energy $E_R$ is measured as $E'$ and incorporates the mean value $\langle E' \rangle = Q_T E_R$, which depends on  the energy dependent quenching factor $Q_T(E_R)$, and the energy resolution $\sigma_{E_R}(E')$. These functions must be measured (although sometimes the energy resolution is just computed).

\begin{figure}[t]
\centering
\includegraphics[width=0.48\textwidth]{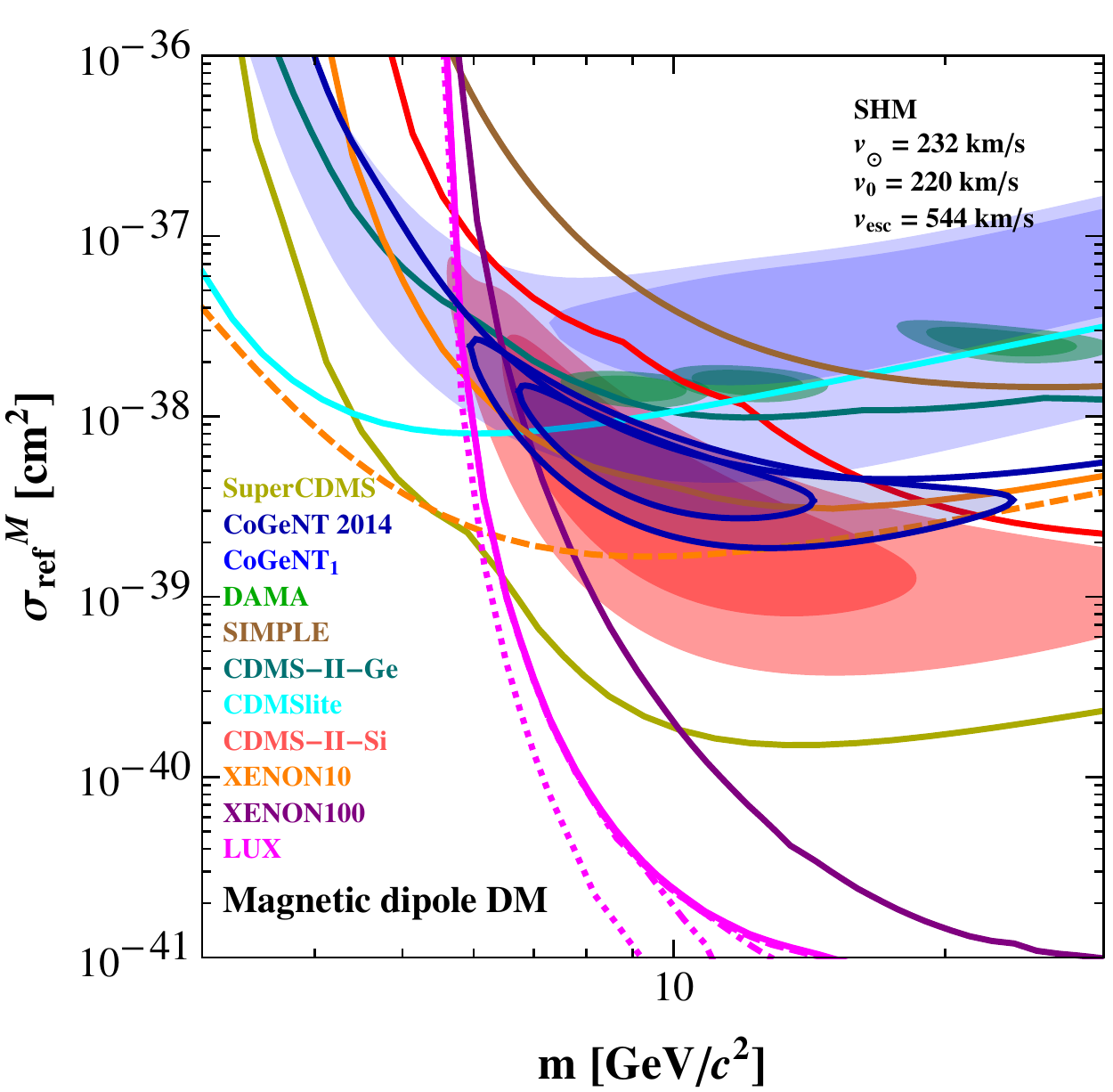}
\label{MDM-Halo-Dep}

\vspace{-0.6cm}

\caption{
$90\%$ CL bounds and $68\%$ and $90\%$ CL allowed regions in the $(m, \sigma_{\rm ref})$ parameter space for MDM, assuming the SHM, with $\sigma_{\rm ref} \equiv \alpha \lambda_\chi^2$. The three  DAMA regions for $Q_{\rm Na} = 0.45$ (left), $Q_{\rm Na} = 0.30$ (middle), and the energy dependent $Q_{\rm Na,Collar}(E_R)$ from. For XENON10 (orange bounds), the solid line is produced by conservatively setting the electron yield $\mathcal{Q}_{\rm y}$ to zero below 1.4 keVnr while the dashed line ignores the $\mathcal{Q}_{\rm y}$ cut. For LUX (magenta bounds), the limits correspond to (from bottom to top) 0, 1, 3, 5, and 24 observed events (see Ref.~\cite{DelNobile:2013gba} for details), however in the range of masses and cross sections depicted here they all overlap apart from the 0 observed event bound. For XENON100 we also show the $68\%$ and $90\%$ CL limits (dashed and dotted line, respectively). Fig. taken from Ref.~\cite{DelNobile:2014eta}.}
\end{figure}

\begin{figure}[t]
\centering
\includegraphics[width=0.49\textwidth]{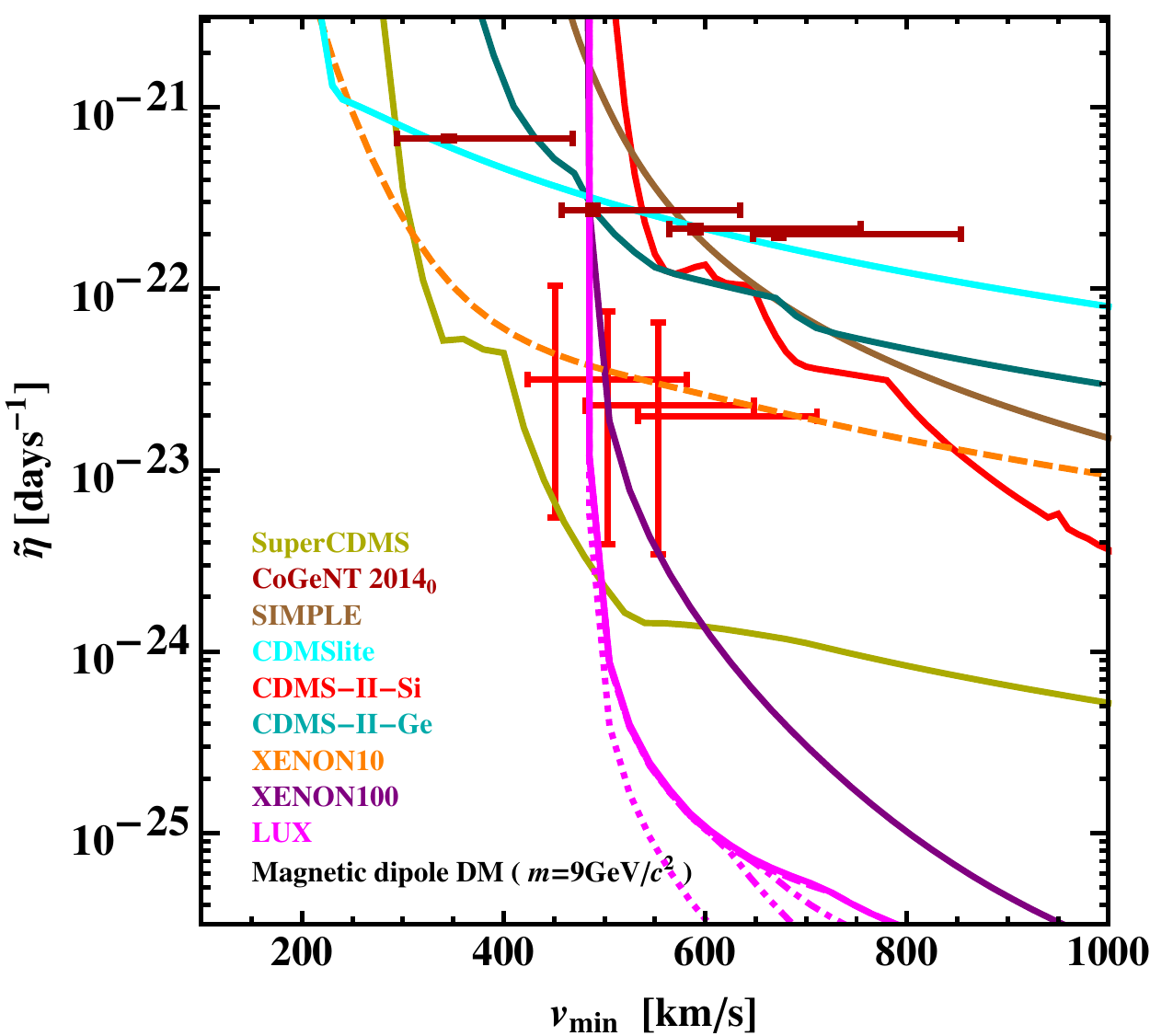}

\vspace{-0.2cm}

\caption{
Measurements of and upper bounds on 
$\overline{\tilde{\eta}^0({\rm v}_{min})} c^2$ and $\overline{\tilde{\eta}^1({\rm v}_{min})} c^2$ for MDM with $m = 9$ GeV/$c^2$. The sodium quenching factor in DAMA is assumed to be $Q_{\rm Na} = 0.30$. For the first and fourth CoGeNT 2014 modulation data points we show the modulus of the negative part of the cross with a thin blue line. The dashed gray lines show the SHM $\tilde{\eta}^0 c^2$ (upper line) and $\tilde{\eta}^1 c^2$ (lower line) for $\sigma_{\rm ref} = 3 \times 10^{-34}$ cm$^2$. Fig. taken from Ref.~\cite{DelNobile:2014eta}.}
\end{figure}

 Statistical analyses usually use rates integrated over energy intervals, e.g. when computing maximum gap limits, 
\begin{eqnarray}
\hspace{-20pt}  R_{[E'_1, E'_2]}(t)& = &\int_{E'_1}^{E'_2} dE' \, \frac{dR}{dE'} =  \frac{\rho}{m} \sum_T \frac{C_T}{m_T} \int_0^\infty dE_R
 \nonumber\\
&  \times & \int_{{\rm v} \geqslant {\rm v}_{min}(E_R)} \hspace{-18pt} d^3 {\rm v} \, f(\vec{{\rm v}}, t) \, {\rm v} \, \frac{d \sigma_T}{d E_R}(E_R, \vec{{\rm v}})
 \nonumber\\
& \times &
 \int_{E'_1}^{E'_2} dE' \, \epsilon(E') G_T(E_R, E') .
 \label{R}
\end{eqnarray}
Changing the order of the $\vec{{\rm v}}$ and $E_R$ integrations in Eq.~\ref{R} we get
\begin{equation}
R_{[E'_1, E'_2]}(t) =  \frac{\rho \sigma_{\rm ref}}{m} \int_0^\infty d^3 {\rm v} \, \frac{f(\vec{{\rm v}}, t)}{{\rm v}} \, \mathcal{H}_{[E'_1, E'_2]}(\vec{{\rm v}}) .
\end{equation}
This relation defines what we call ``integrated response function" $\mathcal{H}_{[E'_1, E'_2]}$. For simplicity, we only consider differential cross sections, and thus $\mathcal{H}_{[E'_1, E'_2]}$ functions, that depend only on the speed ${\rm v}=|\vec{{\rm v}}|$, and not on the whole velocity vector. This is true if the DM flux and the target nuclei are unpolarized and the detection efficiency is isotropic throughout the detector, which is the most common case. With this approximation the detectable integrated rate becomes,
\begin{eqnarray}
\label{R3}
R_{[E'_1, E'_2]}(t) & = & - \int_0^\infty d {\rm v} \, \frac{\partial \tilde{\eta}({\rm v}, t) }{\partial {\rm v}}  \, \mathcal{H}_{[E'_1, E'_2]}({\rm v}) 
 \nonumber\\
 & = &  \int_0^\infty d{\rm v} \, \tilde{\eta}({\rm v}, t) \,  \mathcal{R}_{[E'_1, E'_2]}({\rm v}) ,
\end{eqnarray}
where we have integrated by parts to obtain the second line (the boundary term is zero because the definition of $\mathcal{H}_{[E'_1, E'_2]}({\rm v})$  imposes that  this function is zero  at ${\rm v}=0$) and we have defined a detector and WIMP-nucleus interaction dependent ``response function" as
\begin{eqnarray}
& &  \mathcal{R}_{[E'_1, E'_2]}({\rm v}_{min})
  \equiv  \left. \frac{\partial \mathcal{H}_{[E'_1, E'_2]}({\rm v})}{\partial {\rm v}} \right|_{{\rm v} = {\rm v}_{min}} 
 \nonumber\\
 & = & \sum_T  \frac{4 C_T \mu_T^2}{m^2_T} 
\left\{
\frac{{\rm v}_{min}^3}{\sigma_{\rm ref}} \frac{d \sigma_T}{d E_R}(E_R({\rm v}_{min}), {\rm v}_{min})  \right.
\nonumber\\
& \times & 
\left. \int_{E'_1}^{E'_2} d E' \, \epsilon(E') G_T(E_R({\rm v}_{min}), E') \right. 
\nonumber\\
& + & \left. \int_0^{{\rm v}_{min}} w \, dw  \frac{d}{d {\rm v}_{min}} \left[ \frac{{\rm v}_{min}^2}{\sigma_{\rm ref}} \frac{d \sigma_T}{d E_R}(E_R(w), {\rm v}_{min}) \right] \right.
\nonumber\\
& \times & 
\left.  \int_{E'_1}^{E'_2} dE' \, \epsilon(E') G_T(E_R(w), E')
\right\} .
\end{eqnarray}

For each detected energy interval and particular target nuclide the function $\mathcal{R}_{[E'_1, E'_2]}({\rm v})$ is only nonzero in a certain ${\rm v}_{\min}$ range, as shown in Fig.~\ref{responsefunction}.
Depending on the particular cross section assumed sometimes the response function needs to be regularized to have this property  (see Refs. \cite{DelNobile:2013cva} and \cite{DelNobile:2014eta} for a detailed explanation of the regularization procedure). 

In Ref.~\cite{Gondolo:2012rs} the expression in the last line of Eq. (9) had been derived for SI interactions only, as a generalization of the original formalism~\cite{Fox:2010bz}. The aim of this generalization was to allow the use of efficiencies, energy resolution functions and form factors with arbitrary energy dependence. Fox, Liu, and Weiner~\cite{Fox:2010bz} introduced the halo-independent method for differential  rates and integrated rates, but when integrating the differential rates over  energy bins, took efficiencies and form factors constant over the bin.
 
Using the 2nd. line in Eq.~9 we can map into the $({\rm v}_{min}, \tilde{\eta}^0({\rm v}_{min}))$ parameter space  and the $({\rm v}_{min}, \tilde{\eta}^1({\rm v}_{min}))$ parameter space respectively the measurements and limits on  average integrated rates $R^0_{[E'_1,E'_2]}$ and annual  modulation amplitudes $R^1_{[E'_1,E'_2]}$ of the rates over a detected energy interval $[E'_1,E'_2]$ by different experiments,
\begin{equation}
R_{[E'_1, E'_2]}(t) = R^0_{[E'_1, E'_2]} + R^1_{[E'_1, E'_2]} \cos\!\left[ \omega (t - t_0) \right] ,
\end{equation}
where $t_0$ is the time of the maximum of the signal and $\omega = 2 \pi/$yr. We proceed in the following manner. 

For experiments with putative DM signals $\hat{R}^{\, i}$ in the detected energy ${[E'_1, E'_2]}$  we plot weighted averages of the $\tilde{\eta}^i$ functions with weight $\mathcal{R}_{[E'_1, E'_2]}({\rm v})$,
\begin{equation}
\label{avereta}
\overline{\tilde{\eta}^{\, i}_{[E'_1, E'_2]}} \equiv \frac{\hat{R}^{\, i}_{[E'_1, E'_2]}}
{\int d {\rm v}_{min} \, \mathcal{R}_{[E'_1, E'_2]}({\rm v}_{min})} ,
\end{equation}
 with $i = 0, 1$ for the unmodulated and modulated component, respectively. The interval $[{{\rm v}_{min}}_{,1}, {{\rm v}_{min}}_{,2}]$ in which $\mathcal{R}_{[E'_1, E'_2]}({\rm v}_{min})$ is sufficiently different from zero determines  the width of the ``cross"  in the $({\rm v}_{min}, \tilde{\eta})$ plane corresponding to the $[E'_1, E'_2]$ bin. The vertical bar of the ``cross,'' shows the 1-$\sigma$ error in the rate translated into $\tilde{\eta}$.
 
 To determine the upper bounds on the unmodulated part of $\tilde{\eta}$ set by  experimental upper limit $R^{\rm lim}_{[E'_1, E'_2]}$ on the unmodulated rate in an interval $[E'_1, E'_2]$ (usually at the $90\%$ confidence level)  we follow Refs.~\cite{Fox:2010bz} and \cite{Frandsen:2011gi}: since $\tilde{\eta}^0$ is a non-increasing function of ${\rm v}_{min}$, the smallest possible $\tilde{\eta}^0({\rm v}_{min})$ function passing by a fixed point $({\rm v}_0, \tilde{\eta}_0)$ in the $({\rm v}_{min}, \tilde{\eta})$ plane, is the downward step-function $\tilde{\eta}_0 \, \theta({\rm v}_0 - {\rm v}_{min})$. Thus, assuming the downward step form for $\tilde{\eta}^0({\rm v}_{min})$ we define
 an upper limit at each particular  ${\rm v}_0$ value of ${\rm v}_{min}$
\begin{equation}
\tilde{\eta}^{\rm lim}({\rm v}_0) = \frac{R^{\rm lim}_{[E'_1, E'_2]}}{\int_0^{{\rm v}_0} d {\rm v}_{min} \, \mathcal{R}_{[E'_1, E'_2]}({\rm v}_{min})} .
\end{equation}

In Refs.~\cite{DelNobile:2013cva} and \cite{DelNobile:2014eta} this formalism was applied to MDM, whose differential cross section we presented above in Eq.~\ref{MDMsigma}. Fig.~2 presents the Halo-Dependent comparison of the data, assuming the SHM with parameters given in 
Ref.~\cite{DelNobile:2014eta}. Figs.~3, 4 and 5 present the Halo-Independent data comparison for $m=9$ GeV for the unmodulated part $\tilde{\eta}_0$, the modulated part $\tilde{\eta}_1$, and both parts respectively of the function  $\tilde{\eta}({\rm v}_{min})$, showing all the crosses representing putative measurements and the  most relevant 90\%CL upper bounds.

\begin{figure}[t]
\centering
\includegraphics[width=0.45\textwidth]{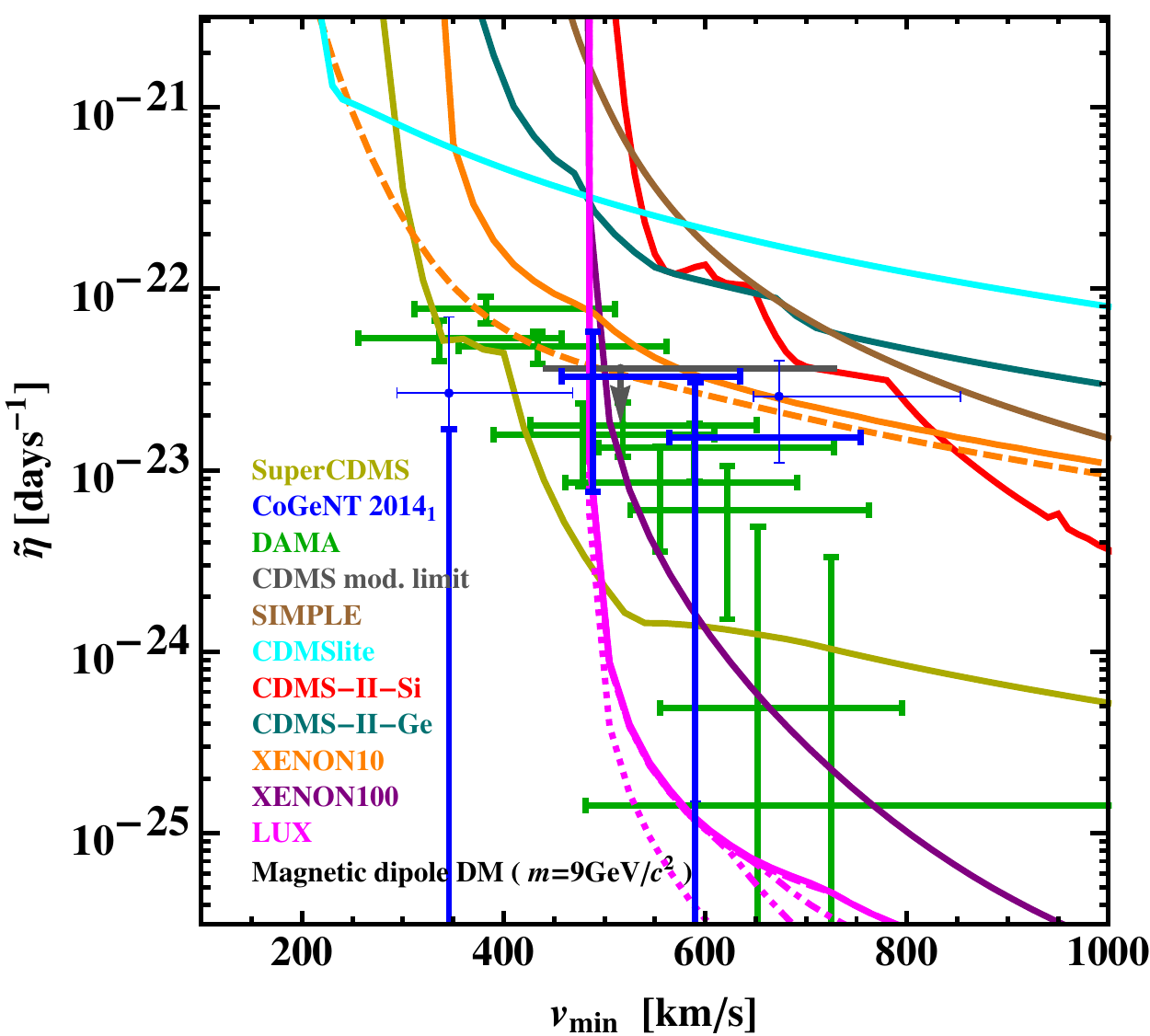}

\vspace{-0.2cm}

\caption{
Same as in Fig.~3 but for  $\tilde{\eta}^1({\rm v}_{min})$.}
\end{figure}

\begin{figure}[t]
\centering
\includegraphics[width=0.45\textwidth]{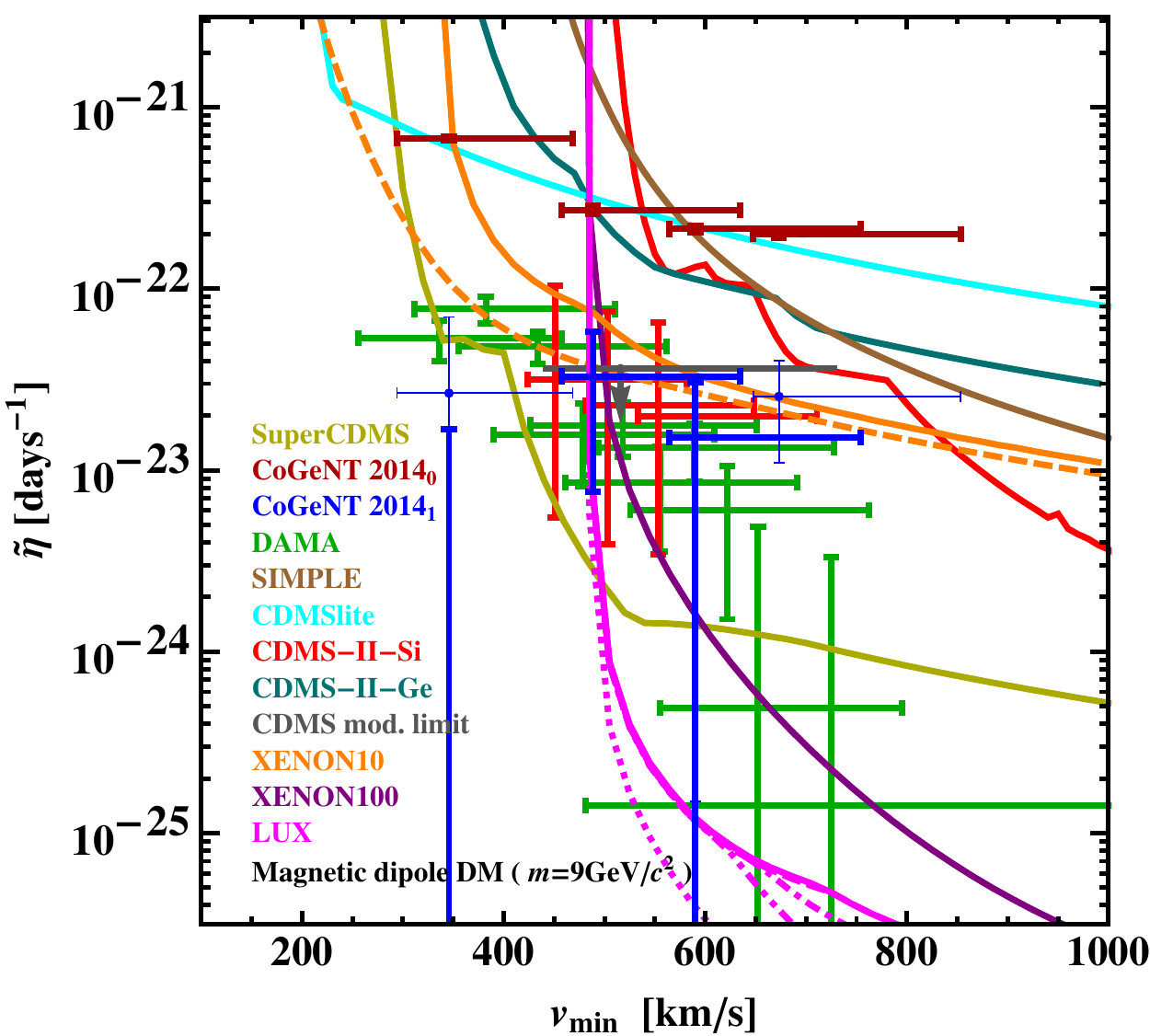}

\vspace{-0.2cm}

\caption{
Same as in Fig.~3 but for both $\tilde{\eta}^0({\rm v}_{min})$  and $\tilde{\eta}^1({\rm v}_{min})$.}
\end{figure}

In the SHM analysis of the allowed regions and bounds in the $(m, \sigma_{\rm ref})$ parameter space (see Fig.~2), CDMSlite, SuperCDMS and LUX set very stringent bounds, and together exclude the allowed regions of three experiments with a positive signal (DAMA, CoGeNT 2011-2012 modulation signal and 2014 unmodulated rate, and CDMS-II-Si) for MDM~\cite{DelNobile:2014eta}. Although in the SHM analysis the DM-signal region is severely constrained by the CDMSlite limit, in the Halo-Independent analysis (presented in Figs.~3 to 5) this limit is much above the DM-signal region~\cite{DelNobile:2013cva, DelNobile:2014eta}. The difference stems from the steepness of the SHM prediction for $\tilde{\eta}^0$
as a function of ${\rm v}_{min}$, which implies that with this halo model $\tilde{\eta}^0$ is constrained at low ${\rm v}_{min}$ by the CDMSlite and other limits.

In the Halo-Independent analysis (see Figs.~3 to 5), although the LUX bound is more constraining than the XENON100 limit, both cover the same range in $v_{min}$ space and are limited to ${\rm v}_{min} \gtrsim 450$ km/s for a WIMP mass of $9$ GeV/$c^2$. This is due to the conservative suppression of the response function below $3.0$ keVnr assumed in this analysis for both LUX and XENON100 (see Ref.~\cite{DelNobile:2013gba} for details). Thus the LUX bound and the previous XENON100 bound exclude mostly the same data for MDM. In other words, almost all the DAMA, CoGeNT (both the 2011-2012 and 2014 data sets), and CDMS-II-Si energy bins that are not excluded by XENON100 are not excluded by LUX either. At lower ${\rm v}_{min}$ values the most stringent bound in this Halo-Independent analysis is the new SuperCDMS limit, which entirely rejects the three CDMS-II-Si crosses. Only the lowest DAMA and CoGeNT modulation data points are not rejected by it.
The situation is of strong tension between the positive and negative direct DM searches results for MDM.

Even without considering the upper limits, in the Halo-Independent analysis of MDM there are problems in the DM signal regions by themselves: as shown in Fig.~5, where the data on $\tilde{\eta}_0$ and $\tilde{\eta}_1$ are overlapped, the crosses representing the unmodulated rate measurements of CDMS-II-Si are either overlapped or below the crosses indicating the modulation amplitude data as measured by CoGeNT (2011-2012 as well as 2014 data sets) and DAMA, which cannot be since the condition  $\tilde{\eta}^1({\rm v}_{min}) < \tilde{\eta}^0({\rm v}_{min})$ must be satisfied (except  possibly at very high ${\rm v}_{min}$, near the speed cutoff).       This indicates strong tension between the CDMS-II-Si data on one side, and DAMA and CoGeNT modulation data on the other (these two seem largely compatible). 

Ref.~\cite{DelNobile:2013cva} has also indicated the way in which the generalized Halo-Independent method presented here should be modified to be able to deal with inelastically scattering DM. In fact,   WIMPs may collide inelastically with  the target nucleus~\cite{TuckerSmith:2001hy}, in which case the initial DM particle scatters to a different state with mass $m' = m + \delta$. This is an interesting possibility which may allow some of the DM hints in direct searches to be compatible with all upper bounds.  DM interacting inelastically via a magnetic dipole moment interaction~\cite{Chang:2010en, Kumar:2011iy} with $\delta >0$, called Magnetic Inelastic DM, MiDM,  may still  allow the DAMA/LIBRA region assumed to be due to DM interactions to be compatible with all negative bounds~\cite{Barello:2014uda}. The mass difference $\delta$ can also be negative, so the inelastic interaction is exothermic~\cite{Graham:2010ca}. It has been recently pointed out that inelastic exothermic DM with Ge-phobic isospin violating interactions could instead make the CDMS-Si region, assumed to be due to DM interactions, compatible  with all  direct searches with negative results, including the SuperCDMS and LUX limits. Both a Halo-Dependent and a Halo Independent  data comparison of direct DM searches for this candidate have been presented in Ref.~\cite{Gelmini:2014psa}. Figs. 6 and 7 present a Halo-Dependent comparison, assuming the SHM, and a Halo Independent comparison, respectively, for Ge-phobic inelastic exothermic DM  taken from Ref.~\cite{Gelmini:2014psa} (see this reference for details).

\begin{figure}[t]
\centering
\includegraphics[width=0.45\textwidth]{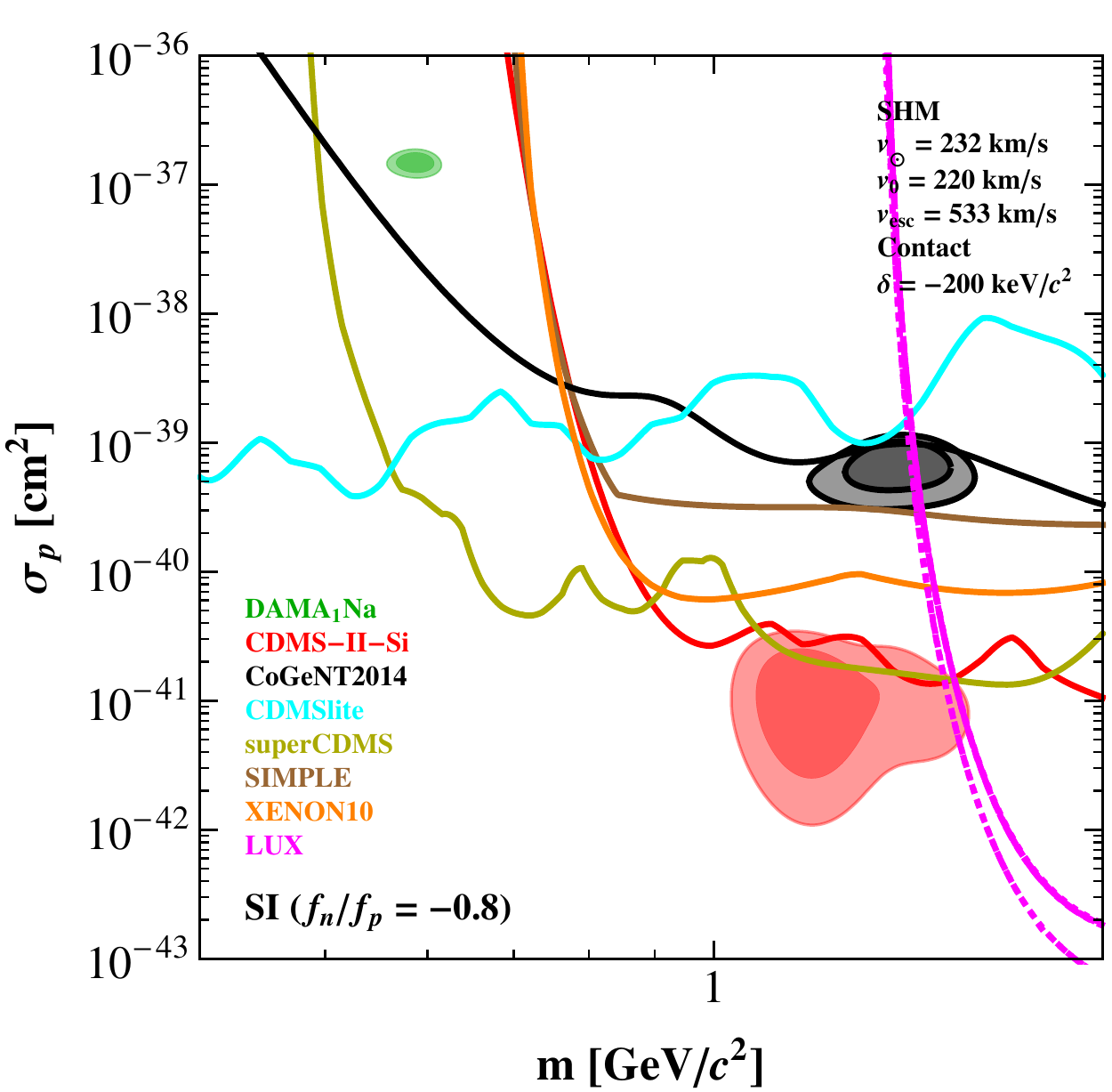}

\vspace{-0.2cm}

\caption{
$90\%$ CL bounds and $68\%$ and $90\%$ CL allowed regions in the WIMP-proton cross section $\sigma_p$ vs WIMP mass plane, assuming the SHM, for spin-independent isospin-violating interactions with $f_n/f_p =-0.8$ (``Ge-phobic"), for inelastic exothermic scattering with $\delta=-200$ keV$/c^2$. See Ref.~\cite{Gelmini:2014psa} for details.}
\end{figure}

\begin{figure}[t]
\centering
\includegraphics[width=0.45\textwidth]{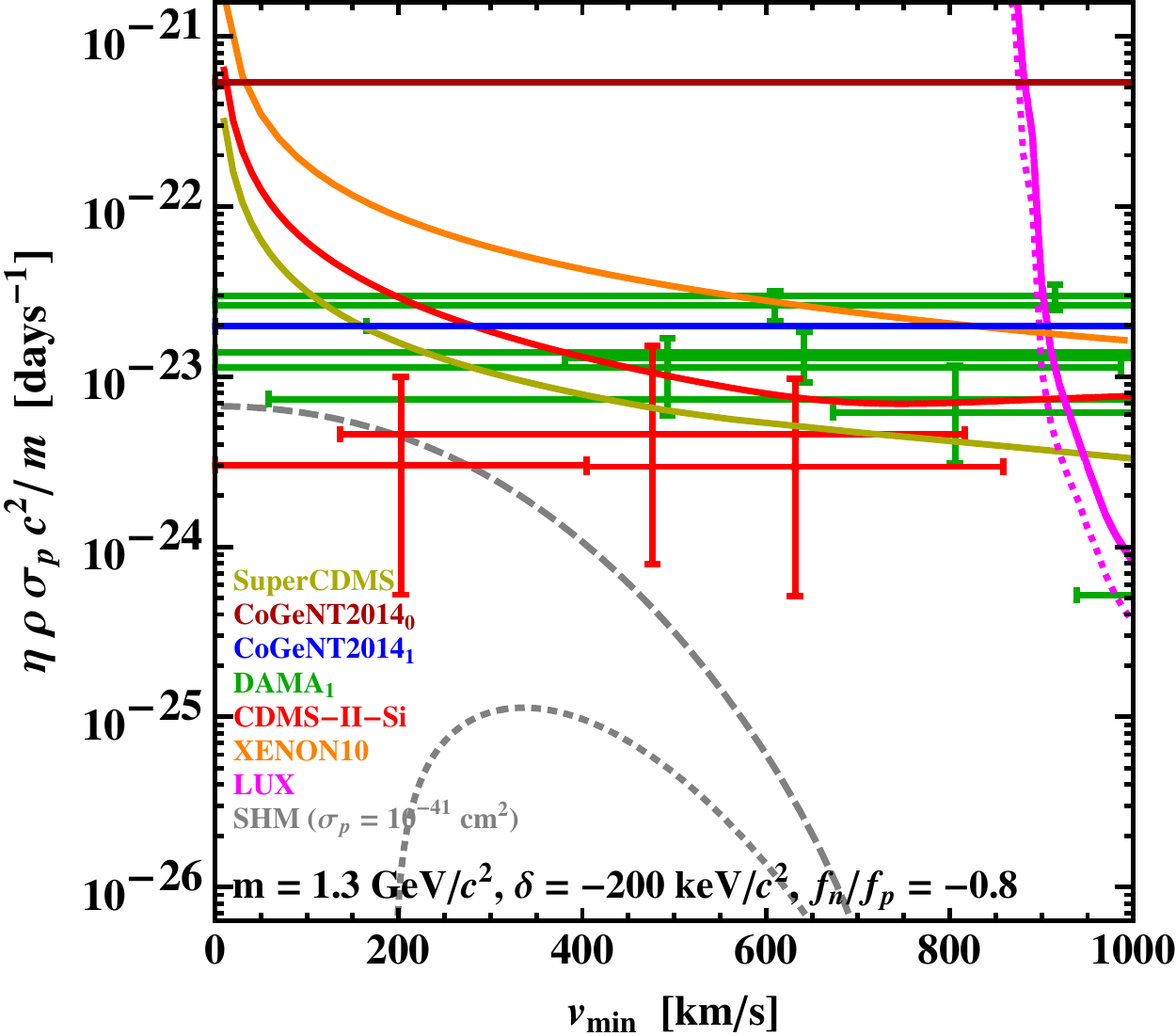}

\vspace{-0.2cm}

\caption{
Measurements of and upper bounds on 
$\tilde\eta^0c^2$  (for CDMS-II-Si and CoGeNT) and $\tilde\eta^1c^2$ (for DAMA) for inelastic exothermic scattering with  $\delta=-200$ keV$/c^2$  for a WIMP with mass $m = 1.3$ GeV/$c^2$  and spin-independent isospin-violating coupling with $f_n/f_p= -0.8$.   The dashed gray lines show the SHM $\tilde{\eta}^0 c^2$ (upper line) and $\tilde{\eta}^1 c^2$ (lower line) for $\sigma_p = 1 \times 10^{-41}$ cm$^2$ which in Fig.~6  is within the CDMS-II-Si region allowed by all upper bounds. See Ref.~\cite{Gelmini:2014psa} for details.
}

\end{figure}

Inelastic DM requires a modification of some of the equations presented above, in particular the definitions of  $\mathcal{H}_{[E'_1, E'_2]}$.  In inelastic scattering, the minimum velocity the DM must have to impart a nuclear recoil energy $E_R$ depends on the mass splitting $\delta$,
\begin{equation}
{\rm v}_{min}= \frac{1}{\sqrt{2 m_T E_R}} \left| \frac{m_T E_R}{\mu_T} + \delta \right| ,
\end{equation}
where $\delta$ can be either positive (endothermic scattering~\cite{TuckerSmith:2001hy}) or negative (exothermic scattering~\cite{Graham:2010ca}) ($\delta = 0$ for elastic scattering). Inverting this equation implies the existence of both a maximum and a minimum recoil energy for a fixed DM velocity ${\rm v}$: $E_R^-({\rm v}) < E_R < E_R^+({\rm v})$, with
\begin{equation}
E_R^\pm({\rm v}) = 
\frac{\mu_T^2 {\rm v}^2}{2 m_T} \left( 1 \pm \sqrt{1 - \frac{2 \delta}{\mu_T {\rm v}^2}} \right)^2 .
\end{equation}
Following the same procedure described above we obtain~\cite{DelNobile:2013cva} a compact form for the integrated response function, 
\begin{eqnarray}
\hspace{-20pt} \mathcal{H}_{[E'_1, E'_2]}(\vec{{\rm v}}) & \equiv &
\sum_T \frac{C_T}{m_T} \int_{E_R^-({\rm v})}^{E_R^+({\rm v})} d E_R \, \frac{{\rm v}^2}{\sigma_{\rm ref}} \frac{d \sigma_T}{d E_R}(E_R, \vec{{\rm v}})
\nonumber\\
& \times &
 \int_{E'_1}^{E'_2} d E' \, \epsilon(E') G_T(E_R, E') .
\end{eqnarray}
The integration limit in the definition of the energy integrated observable rate is not different. Eq.~8 becomes
\begin{equation}
\hspace{-18pt}  R_{[E'_1, E'_2]}(t) =  \frac{\rho \sigma_{\rm ref}}{m} \int_{{\rm v} \geqslant \hat{{\rm v}}_\delta}  d^3 {\rm v} \, \frac{f(\vec{{\rm v}}, t)}{{\rm v}} \, \mathcal{H}_{[E'_1, E'_2]}(\vec{{\rm v}}) ,
\end{equation}
where $\hat{{\rm v}}_\delta$ is the minimum value ${\rm v}_{min}$ can take, 
$\hat{{\rm v}}_\delta = \sqrt{2 \delta / \mu_T}$ for $\delta > 0$ and $\hat{{\rm v}}_\delta = 0$ for $\delta \leqslant 0$
The response function $\mathcal{R}_{[E'_1, E'_2]}$ can then be calculated by taking the partial derivative of 
the integrated response function, as indicated in the first line of Eq.~10. 

As a final comment, let us remark that  the way of comparing direct detection data presented here is not necessarily an inherent  part to the halo independent method but only due to the choice of finding averages over measured energy bins to translate  putative measurements of a DM signal. This may not be the best manner of comparing the direct detection data in $({\rm v}_{min}, \tilde{\eta}({\rm v}_{min}))$ space and more work is necessary to make progress in this respect.

\section*{Acknowledgments}

This talk is based on work done in collaboration with E. Del Nobile,  A. Georgescu, P. Gondolo and Ji-Haeng Hu. G.G. was supported in part by the Department of Energy under Award Number DE-SC0009937.


\begin{thebibliography}{00}
 
 \bibitem{Bernabei:2010mq}
 R.~Bernabei {\it et al.}  [DAMA/LIBRA Coll.],
 Eur.\ Phys.\ J.\  C {\bf 67}, 39 (2010)
 [arXiv:1002.1028 [astro-ph.GA]];
  R.~Bernabei {\it et al.}
  Eur.\ Phys.\ J.\ C {\bf 73}, 2648 (2013)
  [arXiv:1308.5109 [astro-ph.GA]].


\bibitem{Aalseth:2010vx}
 C.~E.~Aalseth {\it et al.}  [CoGeNT collaboration],
 Phys.\ Rev.\ Lett.\  {\bf 106}, 131301 (2011)
 [arXiv:1002.4703 [astro-ph.CO]];
%
 Phys.\ Rev.\ Lett.\  {\bf 107}, 141301 (2011)
 [arXiv:1106.0650 [astro-ph.CO]];
  arXiv:1401.3295 [astro-ph.CO];
  arXiv:1401.6234 [astro-ph.CO].

\bibitem{Agnese:2013rvf} 
  R.~Agnese {\it et al.}  [CDMS Collaboration],
  [arXiv:1304.4279 [hep-ex]].

\bibitem{Angloher:2014myn} 
  G.~Angloher {\it et al.}  [CRESST-II Collaboration],
  arXiv:1407.3146 [astro-ph.CO].


\bibitem{Angloher:2011uu}
 G.~Angloher {\it et al.},
 Eur.\ Phys.\ J.\  C {\bf 72}, 1971 (2012)
 [arXiv:1109.0702 [astro-ph.CO]].

\bibitem{DelNobile:2014sja} 
  E.~Del Nobile, G.~B.~Gelmini, P.~Gondolo and J.~H.~Huh,
  arXiv:1405.5582 [hep-ph]. TAUP 2013 Proceedings.

  \bibitem{Fox:2010bz}
 P.~J.~Fox, J.~Liu and N.~Weiner,
 Phys.\ Rev.\  D {\bf 83}, 103514 (2011)
 [arXiv:1011.1915 [hep-ph]].

\bibitem{Frandsen:2011gi}
 M.~T.~Frandsen {\it et al.}  
 JCAP {\bf 1201}, 024 (2012)
 [arXiv:1111.0292 [hep-ph]].

\bibitem{Gondolo:2012rs}
 P.~Gondolo and G.~B.~Gelmini,
 JCAP {\bf 1212}, 015 (2012)
 [arXiv:1202.6359 [hep-ph]].

\bibitem{Frandsen:2013cna} 
  M.~T.~Frandsen {\it et al.}  
  arXiv:1304.6066 [hep-ph].
  
\bibitem{DelNobile:2013cta} 
  E.~Del Nobile, G.~B.~Gelmini, P.~Gondolo and J.~-H.~Huh,
  arXiv:1304.6183 [hep-ph].
  
\bibitem{HerreroGarcia:2011aa}
 J.~Herrero-Garcia, T.~Schwetz and J.~Zupan,
 JCAP {\bf 1203}, 005 (2012)
 [arXiv:1112.1627 [hep-ph]].
 
\bibitem{HerreroGarcia:2012fu}
 J.~Herrero-Garcia, T.~Schwetz and J.~Zupan,
 Phys.\ Rev.\ Lett.\  {\bf 109}, 141301 (2012)
 [arXiv:1205.0134 [hep-ph]].


\bibitem{Bozorgnia:2013hsa} 
  N.~Bozorgnia, J.~Herrero-Garcia, T.~Schwetz and J.~Zupan,
  JCAP {\bf 1307}, 049 (2013)
  [arXiv:1305.3575 [hep-ph]].
    
\bibitem{DelNobile:2013cva} 
  E.~Del Nobile, G.~Gelmini, P.~Gondolo and J.~H.~Huh,
  JCAP {\bf 1310}, 048 (2013)
  [arXiv:1306.5273 [hep-ph]].
  
\bibitem{DelNobile:2013gba} 
  E.~Del Nobile, G.~B.~Gelmini, P.~Gondolo and J.~H.~Huh,
  JCAP {\bf 1403}, 014 (2014)
  [arXiv:1311.4247 [hep-ph]].
  
\bibitem{DelNobile:2014eta} 
  E.~Del Nobile, G.~B.~Gelmini, P.~Gondolo and J.~H.~Huh,
  JCAP {\bf 1406}, 002 (2014)
  [arXiv:1401.4508 [hep-ph]].
  
\bibitem{Fox:2014kua} 
  P.~J.~Fox, Y.~Kahn and M.~McCullough,
  arXiv:1403.6830 [hep-ph].
  
\bibitem{Gelmini:2014psa} 
  G.~B.~Gelmini, A.~Georgescu and J.~H.~Huh,
  JCAP {\bf 1407}, 028 (2014)
  [arXiv:1404.7484 [hep-ph]].
   
\bibitem{Scopel:2014kba} 
  S.~Scopel and K.~Yoon,
  JCAP {\bf 1408}, 060 (2014)
  [arXiv:1405.0364 [astro-ph.CO]].
  
   
\bibitem{DelNobile:2014sja} 
  E.~Del Nobile, G.~B.~Gelmini, P.~Gondolo and J.~H.~Huh,
  arXiv:1405.5582 [hep-ph].

  
\bibitem{Feldstein:2014ufa} 
  B.~Feldstein and F.~Kahlhoefer,
  arXiv:1409.5446 [hep-ph].
  
\bibitem{Bozorgnia:2014gsa} 
  N.~Bozorgnia and T.~Schwetz,
  arXiv:1410.6160 [astro-ph.CO].
  
  \bibitem{Sigurdson:2004zp}
 K.~Sigurdson {\it et al.} 
 Phys.\ Rev.\  D {\bf 70}, 083501 (2004)
 [Erratum-ibid.\  D {\bf 73}, 089903 (2006)]
 [arXiv:astro-ph/0406355].
%
 V.~Barger, W.~Y.~Keung and D.~Marfatia,
 Phys.\ Lett.\  B {\bf 696}, 74 (2011)
 [arXiv:1007.4345 [hep-ph]].
%
%
 S.~Chang, N.~Weiner and I.~Yavin,
 Phys.\ Rev.\  D {\bf 82}, 125011 (2010)
 [arXiv:1007.4200 [hep-ph]].
%
 W.~S.~Cho et al J.~H.~Huh, I.~W.~Kim, J.~E.~Kim and B.~Kyae,
 Phys.\ Lett.\  B {\bf 687}, 6 (2010)
 [Erratum-ibid.\  B {\bf 694}, 496 (2011)]
 [arXiv:1001.0579 [hep-ph]].
%
 J.~H.~Heo,
 Phys.\ Lett.\  B {\bf 693}, 255 (2010)
 [arXiv:0901.3815 [hep-ph]].
%
 S.~Gardner,
 Phys.\ Rev.\  D {\bf 79}, 055007 (2009)
 [arXiv:0811.0967 [hep-ph]].
%
 E.~Masso, S.~Mohanty and S.~Rao,
 Phys.\ Rev.\  D {\bf 80}, 036009 (2009)
 [arXiv:0906.1979 [hep-ph]].
%
%
 T.~Banks, J.~F.~Fortin and S.~Thomas,
 arXiv:1007.5515 [hep-ph].
%
 J.~F.~Fortin and T.~Tait,
 Phys.\ Rev.\  D {\bf 85}, 063506 (2012)
 [arXiv:1103.3289 [hep-ph]].
%
  K.~Kumar, A.~Menon and T.~M.~P.~Tait,
  JHEP {\bf 1202}, 131 (2012)
  [arXiv:1111.2336 [hep-ph]].
%
 V.~Barger, W.~Keung, D.~Marfatia and P.~Y.~Tseng,
 Phys.\ Lett.\  B {\bf 717}, 219 (2012)
 [arXiv:1206.0640 [hep-ph]].
%
 E.~Del Nobile {\it et al.}
 JCAP {\bf 1208}, 010 (2012)
 [arXiv:1203.6652 [hep-ph]].
%
 J.~M.~Cline, Z.~Liu and W.~Xue,
 Phys.\ Rev.\  D {\bf 85}, 101302 (2012)
 [arXiv:1201.4858 [hep-ph]].
 
\bibitem{Helm:1956zz} 
  R.~H.~Helm,
  Phys.\ Rev.\  {\bf 104}, 1466 (1956).
  
 \bibitem{Pospelov:2008qx}
 M.~Pospelov and A.~Ritz,
 Phys.\ Rev.\  D {\bf 78}, 055003 (2008)
 [arXiv:0803.2251 [hep-ph]].
%
 Y.~Bai and P.~J.~Fox,
 JHEP {\bf 0911}, 052 (2009)
 [arXiv:0909.2900 [hep-ph]].

 \bibitem{TuckerSmith:2001hy}
  D.~Tucker-Smith and N.~Weiner,
  Phys.\ Rev.\ D {\bf 64}, 043502 (2001)
  [hep-ph/0101138].
 
   \bibitem{Chang:2010en}
 S.~Chang, N.~Weiner and I.~Yavin,
 Phys.\ Rev.\  D {\bf 82}, 125011 (2010)
 [arXiv:1007.4200 [hep-ph]].

\bibitem{Kumar:2011iy}
  K.~Kumar, A.~Menon and T.~M.~P.~Tait,
  JHEP {\bf 1202}, 131 (2012)
  [arXiv:1111.2336 [hep-ph]].

\bibitem{Barello:2014uda} 
  G.~Barello, S.~Chang and C.~Newby,
  arXiv:1409.0536 [hep-ph].

\bibitem{Graham:2010ca} 
  P.~W.~Graham, R.~Harnik, S.~Rajendran and P.~Saraswat,
  Phys.\ Rev.\ D {\bf 82}, 063512 (2010)
  [arXiv:1004.0937 [hep-ph]].
  


 \end{thebibliography}
\end{document}